# Improved Charge Transfer Multiplet Method to Simulate M- and L-Edge X-ray Absorption Spectra of Metal-Centered Excited States

**Kaili Zhang[a], Gregory S. Girolami[a] and Josh Vura-Weis[a]***

[a]Department of Chemistry, University of Illinois Urbana-Champaign, Urbana, IL, 61801, USA

Correspondence email: vuraweis@illinois.edu

**Synopsis**  The CTM4XAS software package is extended to simulate the M and L-edge X-ray absorption spectra of metal-centered excited states of 1st-row transition metals. The new capabilities of the method are demonstrated by re-interpreting two previous experimental studies.

**Abstract**  Charge transfer multiplet (CTM) theory is a computationally undemanding and highly mature method for simulating the soft X-ray spectra of first-row transition metal complexes. However, CTM theory has seldom been applied to the simulation of excited state spectra. In this article, we extend the CTM4XAS software package to simulate $M_{2,3}$- and $L_{2,3}$-edge spectra of excited states of first-row transition metals and to interpret CTM eigenfunctions in terms of Russell-Saunders term symbols. We use these new programs to reinterpret the recently reported excited state $M_{2,3}$-edge difference spectra of photogenerated ferrocenium cations and propose alternative assignments for the electronic state of the photogenerated ferrocenium cations supported by CTM theory simulations. We also use these new programs to model the $L_{2,3}$-edge spectra of $Fe^{II}$ compounds during nuclear relaxation following photoinduced spin crossover, and propose spectroscopic signatures for their vibrationally hot states.

**Keywords:**  multiplet simulations; electronic structure; valence excited states; X-ray spectroscopy.

## 1. Introduction

In this paper, we describe several improvements in methods to simulate L- and M-edge spectra of metal-centered excited states of first-row transition metal complexes.  These developments greatly aid in the interpretation of experimental data; for example, they help in deducing such fundamental properties of the excited state as oxidation state, spin state, and coordination geometry. We expect the results to be of use in a variety of X-ray absorption studies that employ soft X-rays and extreme ultraviolet (XUV) light.

Time-resolved spectroscopy has greatly enriched our knowledge of an enormous variety of important and fundamental photophysical processes (Zewail, 2000). Although much of the work in this area has involved the use of light at IR, visible, and UV energies, additional information of great value can be obtained by means of time-resolved experiments at soft X-ray energies (Bressler & Chergui, 2004, 2010; Chen, 2005; Chen *et al.*, 2014; Milne *et al.*, 2014). Short pulses of X-ray photons with energies on the order of 400 - 900 eV are available for time-resolved spectroscopic experiments at synchrotron and free-electron laser facilities. Photons within this energy range, which are able to promote 2p-to-valence ($L_{2,3}$-edge) transitions of first-row transition metals, have been fruitfully applied in studying many processes, such as the ultrafast energy dissipation of iron(II) photosensitizers (Huse *et al.*, 2010; Cho *et al.*, 2012; Hong *et al.*, 2015).

The analogous 3p-to-valence ($M_{2,3}$-edge) transitions of first-row transition metals can be probed by XUV photons with energies on the order of 30 - 80 eV. Recently, femtosecond and even attosecond XUV pulses of such photons have become available through high-harmonic generation (HHG) (Baker *et al.*, 2014; Vura-Weis *et al.*, 2013; Goulielmakis *et al.*, 2010; Chatterley *et al.*, 2016; Jiang *et al.*, 2014), thus enabling core-level studies of chemical and physical processes at femtosecond time resolutions by means of table-top instrumentation. The relative convenience of an HHG light source compared to a synchrotron or a free-electron laser promises to make time-resolved soft X-ray studies more readily available.[1]

The increased use of time-resolved soft X-ray and XUV methods has created a need for better theoretical tools, especially to carry out spectroscopic simulations. An ideal tool would be computationally undemanding while still capable of explaining pertinent spectroscopic features. Soft X-ray spectra are heavily dominated by multiplet effects stemming from strong p-d interactions, and thus are difficult to predict using methods based on single particle models (de Groot, 2005; de Groot & Kotani, 2008). Time dependent density functional theory (TD-DFT) in its standard form is incapable of capturing the interplay of electron-electron coulombic interaction and spin-orbit coupling relevant to a 2p/3p core-hole in the presence of unpaired valence electrons (Milne *et al.*, 2014; Josefsson *et al.*, 2012).

---

[1] Certain related techniques have the potential of further expanding the scope of time-resolved soft X-ray spectroscopy. For example, $K_\beta$ XES and 1s2p RIXS enable time-resolved soft X-ray to be performed on reactors in operando (Milne *et al.*, 2014; de Groot *et al.*, 2005; Zhang *et al.*, 2014) and electron energy loss spectroscopy (EELS) enables time-resolved soft X-ray spectroscopy to be performed on nano-sized structures (van der Veen *et al.*, 2015).

Several novel quantum chemical methods have been developed to model multiplet effects from first principles (Milne *et al.*, 2014). Among the most successful of these are RASSCF (Josefsson *et al.*, 2012) and DFT/ROCIS (Roemelt *et al.*, 2013). Both methods take advantage of configuration interaction, and, consequently, scale as $O(N^5)$ (Roemelt *et al.*, 2013). Compared to TD-DFT, both RASSCF and DFT/ROCIS capture more completely the p-d interactions and spin-orbit coupling involved in soft X-ray spectroscopy (Milne *et al.*, 2014). Although RASSCF and DFT/ROCIS eschew explicit parametrization of electronic structure, both methods contain a minimal but still critical *ad hoc* element, namely, the choice of active spaces in RASSCF and the choice of the underlying functional in DFT/ROCIS, necessitating trial-and-error tuning for accurate simulations (Milne *et al.*, 2014).

The traditional approach to deal with multiplet effects has been semi-empirical charge transfer multiplet (CTM) theory (de Groot & Kotani, 2008). CTM theory, which is based on atomic multiplet theory (Cowan, 1981), models electron-nuclear interactions, electron-electron interactions, and spin-orbit coupling with a parametric Hamiltonian (de Groot, 2005; de Groot & Kotani, 2008), Ligands are modeled as an electrostatic crystal field conforming to some predetermined point group, although covalency in metal-ligand bonding can be treated with additional parameters. By adjusting the various parameters, good agreement between simulation and experiment can be achieved, making CTM theory an excellent method for analyzing experimental spectra (Roemelt *et al.*, 2013). For example, semi-empirical CTM theory has been highly successful in simulating the experimental soft X-ray spectra of first-row transition metal complexes, including those with unpaired electrons and orbital angular momentum (de Groot & Kotani, 2008; de Groot, 2005). The systematic variation of parameters within a series of related compounds can be analyzed to afford useful physical insights (Hocking *et al.*, 2006, 2007).

Compared with quantum chemical methods such as RASSCF and DFT/ROCIS, semi-empirical CTM theory has weaker predictive power owing to its extensive parametrization, and less flexibility owing to its use of group theoretic crystal field parameters instead of real space atomic coordinates. On the other hand, semi-empirical CTM theory is significantly less demanding computationally than quantum chemical methods: typical CTM theory calculations require only a matter of minutes on single-core desktop computers.

Its low computational demand, and its demonstrated ability to give realistic simulations of soft X-ray spectra, recommend CTM theory both as an excellent tool of first resort for exploring unknown systems and as a method for analyzing experimental spectra, especially of complexes with high symmetry and with metal-dominated electronic structures. However, to date, CTM theory has seldom been used to simulate soft X-ray (or XUV) spectra of d-d excited states (Vura-Weis *et al.*, 2013).

In this paper, we demonstrate that CTM theory methods can be used to simulate $L_{2,3}$- and $M_{2,3}$-edge spectra of metal-centered excited states of first-row transition metal complexes. In the first of two case studies, we reanalyze the $M_{2,3}$-edge spectra of ferrocenium cations generated by strong-field ionization (Chatterley *et al.*, 2016) and propose a new assignment of the electronic state responsible for the spectroscopic features. In the second case study, we explore the evolution of the Fe $L_{2,3}$-edge spectrum of $Fe^{II}$ polypyridyl complexes during nuclear relaxation following photoinduced spin crossover (Huse *et al.*, 2010). We describe the spectroscopic changes that CTM theory predicts should occur as the metal center relaxes from a Frank-Condon state to a metastable state.

## 2. Methods

### 2.1. Computation of simulated spectra

The atomic structure code of Cowan (Cowan, 1981), the group theory program of Butler (Butler, 1981), and the CTM theory program of Kotani and Thole (Thole *et al.*, 1985), all supplied as part of the CTM4XAS 5.5 package (Stavitski & de Groot, 2010), were used to compute the eigenstates of the parametric Hamiltonian and the stick spectra of L- or M-edge excitations of all d-d excited states. For L-edge spectra, the $L_3$-edge and $L_2$-edge sticks are broadened with Voigt profiles having Lorentzian FWHMs ($\Gamma$) of 0.2 and 0.4 eV, respectively, and a Gaussian width ($\sigma$) of 0.2 eV (Hocking *et al.*, 2006). For M-edge spectra, the sticks are broadened with asymmetric Fano line shapes with a Fano asymmetry parameter ($q$) of 3.5 (Fano, 1961; Vura-Weis *et al.*, 2013). Owing to the term-dependent variability of the lifetimes of 3p core-holes, $\Gamma$ is computed for each 3p-3d transition by using a modified version of the Auger program of Kotani and Thole (de Groot & Kotani, 2008; Okada & Kotani, 1993) (source code in Supporting Information) whereas $\sigma$ is kept constant at 0.2 eV (Zhang *et al.*, 2016). A Python program was written to streamline the process of spectrum computation and plotting (source code in Supporting Information, see section S2 of Supporting Information for details).

### 2.2. Assignment of CTM theory eigenstates

Because valence-level spin-orbit coupling in first-row transition metals is weak, valence excited states of first-row transition metal complexes are usually described by spin-orbit uncoupled Russell-Saunders term symbols; each state is identified by its spin quantum number and its irreducible representation in the point group of the ligand field. In contrast, for computational efficiency, CTM4XAS calculates the eigenfunctions of the parametric Hamiltonian in a spin-orbit coupled basis (Laan, 2006). Consequently, each state is identified in the CTM4XAS output only by a spin-orbit-coupled irreducible representation. In order to facilitate the identification of d-d excited states, we must transform the CTM4XAS-provided eigenfunctions into uncoupled basis functions using a generalized form of Clebsch-Gordan coefficients (Butler, 1981; Piepho & Schatz, 1983).

For a more concrete example, consider a calculation in the octahedral point group O. The spin-orbit coupled basis functions are labelled as $|(SL)Ja_J\Gamma_O^J\rangle$, where $S$, $L$ and $J$ are the free-ion spin angular momentum, orbital angular momentum, and total angular momentum quantum numbers, $a_J$ is the branching multiplicity index from SO(3) to O, and $\Gamma_O^J$ is the irreducible representation in O. An eigenfunction $|\Psi\rangle$ of an ion with $N$ d-electrons in an octahedral environment is then expressed as $|\Psi\rangle = \sum_{(SL)Ja_J\Gamma_O^J} A_{(SL)Ja_J\Gamma_O^J} |(SL)Ja_J\Gamma_O^J\rangle$, where the summation ranges over all combinations of indices allowed for a $d^N$ system. A spin-orbit coupled basis function $|(SL)Ja_J\Gamma_O^J\rangle$ can be written as a linear combination of spin-orbit decoupled basis functions $|(SL)Ja_J\Gamma_O^J\rangle = \sum_{La_L\Gamma_{Oh}^L, Sa_S\Gamma_O^S} \langle La_L\Gamma_O^L, Sa_S\Gamma_O^S | (SL)Ja_J\Gamma_O^J\rangle_r |(La_L\Gamma_O^L, Sa_S\Gamma_O^S)r\Gamma_O^J\rangle$, where $a_L$ is the orbital branching index, $\Gamma_O^L$ is the orbital irreducible representation, and, analogously for the spin indices, $r$ is a product multiplicity index in case $\Gamma_O^J$ appears multiple times in the direct product of $\Gamma_{Oh}^L$ and $\Gamma_{Oh}^S$. The coupling coefficients $\langle La_L\Gamma_O^L, Sa_S\Gamma_O^S | (SL)Ja_J\Gamma_O^J\rangle_r$, being intrinsic properties of the groups SO(3) and O, are independent of the identity of the ion being simulated. Therefore, a large set of coupling coefficients sufficient for the decomposition all $d^N$ and $3p^5 3d^N$ systems can be precomputed. Tabulated values of these coefficients for various pairs of groups have been published (Butler, 1981; Piepho & Schatz, 1983). Using these values, any spin-orbit coupled eigenfunction can then be expressed in the decoupled basis $|\Psi\rangle = \sum_{La_L\Gamma_O^L, Sa_S\Gamma_O^S} A'_{La_L\Gamma_O^L Sa_S\Gamma_O^S r} |(La_L\Gamma_O^L, Sa_S\Gamma_O^S)r\Gamma_O^J\rangle$, where $A'_{La_L\Gamma_O^L, Sa_S\Gamma_O^S} = \sum_{(SL)Ja_J\Gamma_O^J} A_{(SL)Ja_J\Gamma_O^J} \langle La_L\Gamma_O^L, Sa_S\Gamma_O^S | (SL)Ja_J\Gamma_O^J\rangle_r$. The make-up of $|\Psi\rangle$ in terms of pure-spin Russell-Saunders terms can then be determined by examining the values of $\left|A'_{La_L\Gamma_O^L, Sa_S\Gamma_O^S}\right|^2$ for different combinations of $L$, $S$, and $\Gamma_O^L$. In the special case where spin-orbit coupling is set to zero, the basis functions contributing to an eigenfunction $|\Psi\rangle$ will have identical values for $\Gamma_O^L$ and $S$.

The algorithm for the basis transformation was implemented in a Python program (source code and accompanying data files in Supporting Information, see section S1 of Supporting Information for details).

## 3. Results and discussion

### 3.1. Ferrocene and photogenerated ferrocenium ions

Recently, Chatterley et al. reported the $M_{2,3}$-edge spectrum of gas phase ferrocenium cations produced by the strong-field photoionization of ferrocene vapor (Chatterley *et al.*, 2016). The authors simulated the $M_{2,3}$-edge spectra of various possible ground and excited states by restricted energy window TDDFT (REW-TDDFT) based on the B3LYP functional.

Before proceeding to simulations of the spectra of photogenerated ferrocenium ions, we first used extended CTM4XAS to simulate the spectrum of neutral ferrocene in its ground state, based on ligand field parameters given by Gray et al (Gray *et al.*, 1971). An empirical, uniform horizontal shift was applied to the computed spectrum, in order to correct for known inaccuracies in the absolute transition energies predicted by CTM theory (Vura-Weis *et al.*, 2013); here the shift was 3.7 eV, chosen to match the peak at 59 eV in the experimental spectrum. Relative to the REW-TDDFT spectrum, the CTM theory simulation (Figure 1) more closely reproduces the general two-peak structure of the observed spectrum, except for the low-energy shoulder at 57 eV, which the CTM theory simulation lacks. Examination of the computed CTM stick spectrum shows that the large peak at 59.5 eV is composed of a distribution of transitions, several of which – particularly those at 58.2 eV and 57.4 eV – are at much lower energies than the rest. CTM theory may have underestimated the intensities of some of these lower-energy transitions, causing them to merge into the large peak at 59.5 eV.

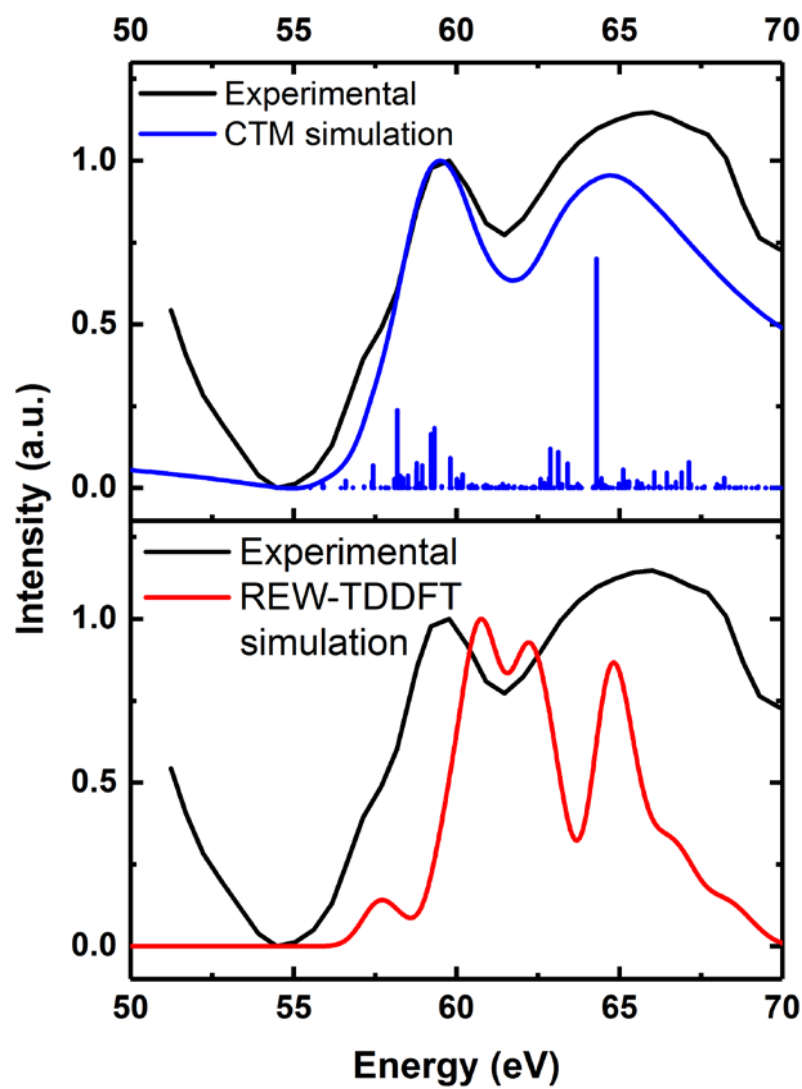

**Figure 1** $M_{2,3}$-edge absorption spectra of the ground state of neutral ferrocene: experimental, CTM simulation and REW-TDDFT simulation (Chatterley *et al.*, 2016).

As described by Chatterley et al., the difference spectrum (excited state minus ground state) of the photogenerated ferrocenium cation contains two positive difference peaks at 53.2 eV and 55.7 eV, both of which are lower in energy than the resonant absorption features of neutral ferrocene; these peaks lie in a region where the spectrum of ferrocene is relatively smooth and featureless (Figure 2) (Chatterley *et al.*, 2016). Previous authors simulated the $M_{2,3}$-edge spectra of various multiplet states of ferrocenium by REW-TDDFT. Through a comparison of the low energy (50 – 56 eV) portion of the simulated absorption spectra to the observed difference spectrum, they concluded that the photogenerated ferrocenium cations are mixture of species, some in the $^2A_1$ state and the others in the $^4E_2$ state.

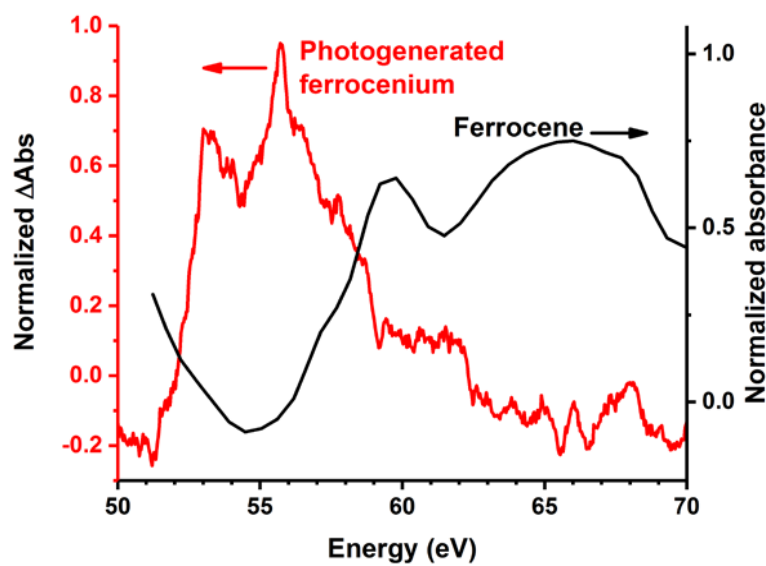

**Figure 2** $M_{2,3}$-edge difference spectrum of photogenerated ferrocenium superimposed on the absorption spectrum of ferrocene (Chatterley *et al.*, 2016).

We used extended CTM4XAS with crystal field parameters taken from a previous spectroscopic study (Gray *et al.*, 1971) to compute the absorption spectra of various excited states of the ferrocenium cation. The 3d-3d Slater-Condon parameters were varied from 57% to 100% of the free-ion values in order to find a best fit value. The proportion by which the Slater-Condon parameters are reduced, which reflects the extent of delocalization of metal-based electron density onto ligand-based orbitals, is known as the nephelauxetic factor. The CTM theory eigenstates were analyzed by decomposition into spin-orbit decoupled basis functions; Table 1 gives these decompositions for one value of the nephelauxetic factor, 86% (tables for other nephelauxetic factors are included in section S1.4 of Supporting Information). Decomposition analysis showed that each Russell-Saunders term is split by spin-orbit coupling into several component eigenfunctions separated by no more than 0.1 eV. In principle, the excited state of the photogenerated ferrocenium cation exists in a linear combination of the component eigenfunctions. However, due to the complexities of the strong-field ionization process, the relative phases of the component functions cannot be readily determined. For convenience, the lowest energy eigenfunction of each Russell-Saunders term is chosen as the representative for simulation.

**Table 1** Assignment of excited states of ferrocenium cation with 86% Slater-Condon scaling.

| Energy | $\Gamma^J$, order within $\Gamma^J$** | Purity | Assignment |
|---|---|---|---|
| -5.33* | $\frac{1}{2}$, 1 | 95.7% | $^6A_{1g}$ |
| -5.34 | $\frac{3}{2}, 2$ | 92.7% | |
| -5.35 | $\frac{5}{2}, 2$ | 86.9% | |
| -5.06* | $\frac{1}{2}, 3$ | 96.3% | |
| -5.00 | $\frac{1}{2}, 4$ | 73.4% | ${}^4E_{1g}$ |
| -5.08 | $\frac{3}{2}, 3$ | 92.4% | |
| -5.10 | $\frac{5}{2}, 3$ | 86.4% | |
| -4.90* | $\frac{1}{2}, 5$ | 100% | |
| -4.88 | $\frac{3}{2}, 4$ | 99.8% | ${}^4E_{2g}$ |
| -4.84 | $\frac{3}{2}, 5$ | 99.7% | |
| -4.86 | $\frac{5}{2}, 4$ | 99.7% | |
| -5.12* | $\frac{1}{2}, 2$ | 72.1% | ${}^2A_{1g}$ |
| -6.10* | $\frac{3}{2}, 1$ | 99.4% | ${}^2E_{2g}$ |
| -6.22 | $\frac{5}{2}, 1$ | 99.2% | |

* Chosen as representative for simulation.

** Irreducible representations $\Gamma^J$ are given in Butler notation (Butler, 1981). The entries for $\Gamma^J = -\frac{5}{2}$, being identical to entries for $\Gamma^J = \frac{5}{2}$ of the corresponding terms, have been omitted.

Simulated difference spectra were computed by subtracting the simulated absorption spectrum of neutral ferrocene from the simulated absorption spectrum of each excited state. A uniform horizontal shift of 1.6 eV has been applied to each of the spectra of ferrocenium excited states before subtraction.

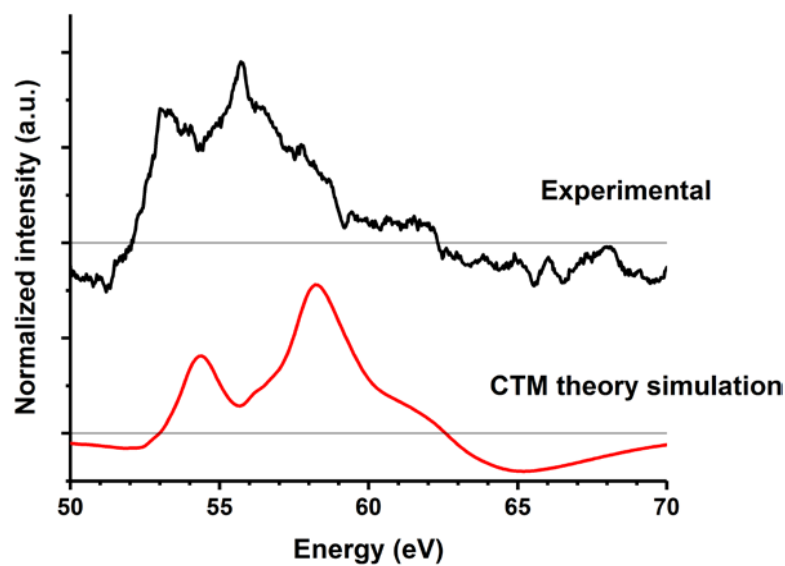

**Figure 3** Experimental difference spectrum of photogenerated ferrocenium and CTM theory simulation of $^6A_1$ state of ferrocenium (Chatterley *et al.*, 2016).

The best qualitative match to the experimental spectrum is achieved with the $^6A_1$ state of ferrocenium, in which the nephelauxetic factor is 86% (Figure 3). The simulated difference spectra of other states show several undulatory features above 60 eV that are not observed in the experimental spectrum (Figure 4). The simulated spectrum accurately reproduces the two-peak structure of the experimental spectrum, and also contains shoulders that may also be present in the latter. The calculated inter-peak spacing (3.8 eV) is larger than the observed spacing (2.5 eV), which suggests that the 3p-3d electron-electron interaction is overestimated by the atomic Hartree-Fock algorithm that underlies CTM4XAS. (See section S3 of Supporting Information for a brief exploration of the effects of 3p-3d interactions on the spectra.) The simulated difference features are also blueshifted by 1-2 eV relative to the observed difference features.[2]

---

[2] With the simulated ground state spectrum fixed, redshifting the simulated excited state spectrum leads to further overestimation of the inter-peak spacing without appreciably affecting the position of the peak at 58 eV in the difference spectrum.

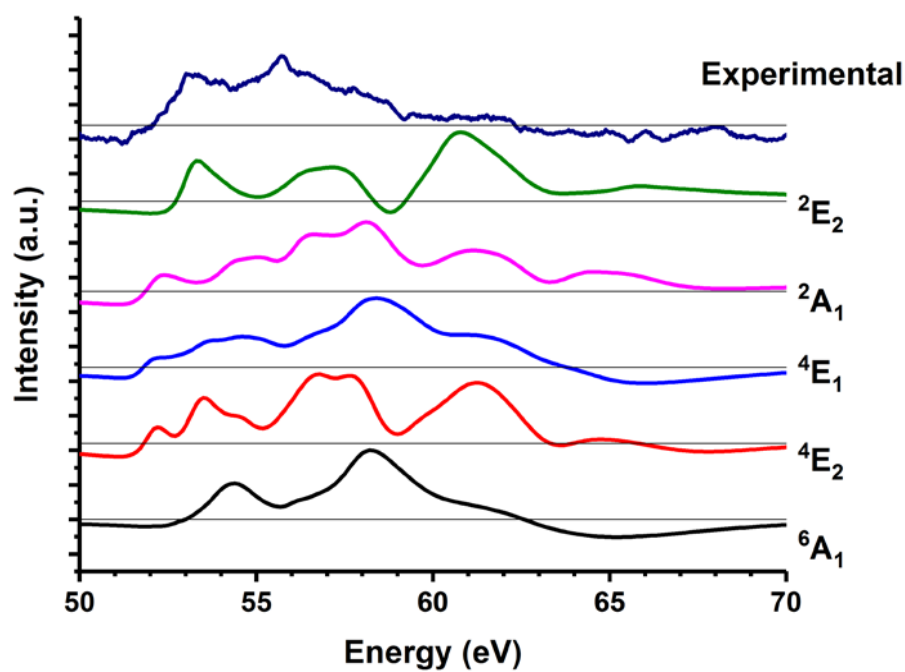

**Figure 4** Simulated difference spectra of various multiplet excited states of ferrocenium (Chatterley *et al.*, 2016).

Our best-fit nephelauxetic factor of 86% differs from the values of 42% or 74% suggested from studies of ferrocenium salts[3] (Gray *et al.*, 1971) (see Section S4 of Supporting Information for a brief exploration of the effects of 3d-3d interactions on the $^6A_1$ difference spectrum). If the nephelauxetic factor is 42%, the $^6A_1$ state sits at least 1.5 eV above the ground state, whereas for a larger value of 86%, the $^6A_1$ state is only 0.8 eV above the ground state due to the increased electron-electron repulsion favoring high-spin configurations (Figure 5). It needs to be noted that the nephelauxetic parameter was not particularly well constrained by the available UV/Vis data (Gray *et al.*, 1971). A nephelauxetic parameter of 86% is supported by the reported experimental $L_{2,3}$-edge spectrum of ferrocenium hexafluorophosphate (see Section S5 of Supporting Information) (Otero *et al.*, 2009).

---

[3] The free-ion Slater-Condon parameters computed by CTM4XAS are $F^2_{dd}$ = 11.0 eV and $F^4_{dd}$ = 6.82 eV, corresponding to a Racah B parameter value of $B_{freeion}$ = 0.117 eV (945 cm$^{-1}$) (Cowan, 1981). The Racah B parameter values given by Gray et al., 390 cm$^{-1}$ or 700 cm$^{-1}$, corresponds to nephelauxetic factors of 42% or 74%, respectively.

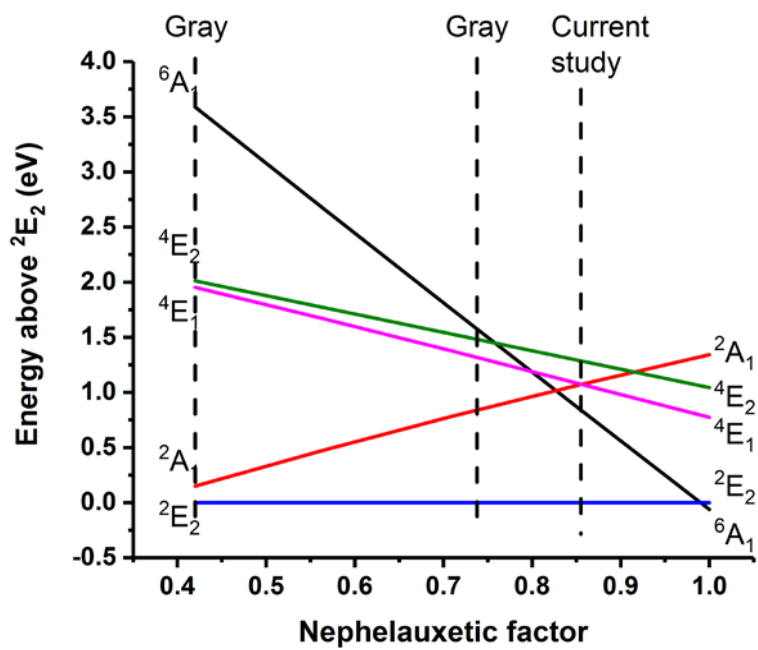

**Figure 5** Excited state energies of ferrocenium as a function of nephelauxetic factor as computed by CTM theory.

This case study convincingly demonstrates that CTM theory simulations are more in agreement with the experimental spectra of metal-centered excited states of the ferrocenium cation than REW-TDDFT simulations. Furthermore, the CTM simulations suggest that the photogenerated ferrocenium cation in this experiment is in a vibrationally and/or electronically excited $^6A_1$ state that has significant free-ion character.

### 3.2. Photoinduced spin transition in Fe$^{II}$

In this second case study, we examine the ability of extended CTM4XAS to simulate the soft X-ray spectra of metal centered excited states in which the metal centers change from low spin (LS) to high spin (HS) following photoexcitation.

Polypyridyl Fe$^{II}$ complexes such as $Fe(bpy)_3^{2+}$ and $Fe[Tren(Py)_3]^{2+}$ are known to exhibit ultrafast relaxation into metastable quintet excited states following excitation of the MLCT band of their singlet ground states (Cho *et al.*, 2012). Transient XANES spectra at the Fe K-edge suggested that the spin crossover to the quintet electronic state from the MLCT state occurs with 100% quantum yield and with a time constant of less than 150 fs; this time scale corresponds to only two times the oscillation period of the ligand cage breathing mode (Bressler & Chergui, 2010; Zhang *et al.*, 2014). Recent UV/Vis pump-probe experiments with 40 fs time resolution suggested that the generation of the quintet state may well be complete in less than 50 fs, with all subsequent spectroscopic changes being ascribable to vibrational relaxation (Auböck & Chergui, 2015). More recent transient hard X-ray studies with a 30 fs time resolution performed using a free-electron laser light source suggested that the quintet species undergoes significant geometrical relaxation after spin-crossover (Lemke *et al.*, 2017).

Ultrafast soft X-ray spectroscopy probes strong, dipole-allowed transitions into metal centered orbitals. This technique can give insights into the evolution of the metal centered electronic and geometric structures as the metal center relaxes into its metastable excited state. The transient $L_{2,3}$-edge spectra of the quintet metastable state of a number of $Fe^{II}$ polypyridyl complexes have been reported (Huse *et al.*, 2010; Cho *et al.*, 2012). In all cases, with the time-resolution available, the photogenerated quintet species relaxes sufficiently quickly that its spectrum can be simulated as the ground state of a HS model system with a reduced ligand field (Figure 6) (Monat & McCusker, 2000; Consani *et al.*, 2009).

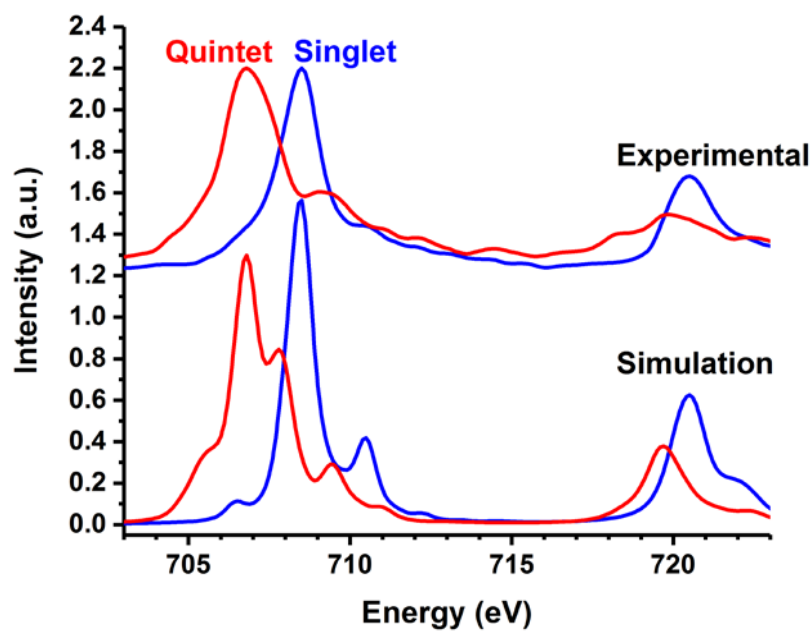

**Figure 6** Top: Experimental $L_{2,3}$-edge spectra of Fe[Tren(Py)$_3$](PF$_6$)$_2$ in the ground singlet state (blue) and in the photogenerated quintet state (red). Bottom: LFM simulations of an $Fe^{II}$ cation in $^1A_1$ state (blue) with 10Dq = 2.2 eV and in $^5T_2$ state (red) with 10Dq = 0.6 eV (Huse *et al.*, 2010).

However, before being vibrationally cooled into the metastable state, the Frank-Condon photogenerated quintet species remains a bona fide excited state and cannot be approximated as the ground state of a model system. The experimental spectra of such excited Frank-Condon states formed early in the relaxation process will increasingly become available with improvements in the time resolution at free-electron laser instruments (Lemke *et al.*, 2017). The extensions to CTM4XAS described herein allow the evolution of the Fe $L_{2,3}$-edge spectrum to be predicted over the entirety of the relaxation process. An approach to do so follows.

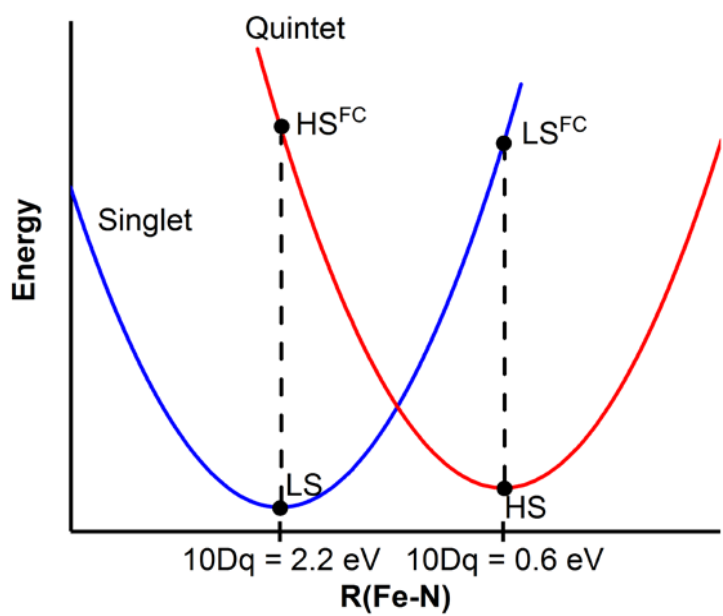

**Figure 7** Schematic potential energy surface relevant to $Fe^{II}$ spin-crossover complexes.

In Figure 7, the Franck-Condon quintet state is labelled $HS^{FC}$, whereas the relaxed geometry state is labelled HS. Although the $HS^{FC}$ state is accessed indirectly through the MLCT manifold, the rapidity of the process means that the geometry of the $HS^{FC}$ state can be approximated as being identical to the LS geometry.[4] For simplicity, only relaxation along the primary reaction coordinate, the symmetric Fe-N stretching mode, will be considered (Lemke *et al.*, 2017). Because CTM theory models the effect of ligands as an electrostatic crystal field, the octahedral crystal field strength, 10Dq, will be used as a proxy for measuring geometric distortion. Following previous authors, 10Dq values of 2.2 eV and 0.6 eV are used for the low-spin and high-spin environments, respectively (Huse *et al.*, 2010). Similarly to the previous section, the CTM theory eigenfunctions are analyzed by decomposition into spin-orbit decoupled basis functions. The $^5T_{2g}$ state is split by the interaction of spin-orbit coupling and octahedral crystal field into six components spaced by less than 0.1 eV (Figure 8). The lowest energy component is chosen as the representative for simulation.

---

[4] Analogously, a continuum of excited singlet states ranging from the Frank-Condon singlet species ($LS^{FC}$, Figure 7) to the relaxed singlet species (LS, Figure 7) is involved in reverse-light induced spin state trapping (reverse-LIESST) (Hauser, 1986, 2004).

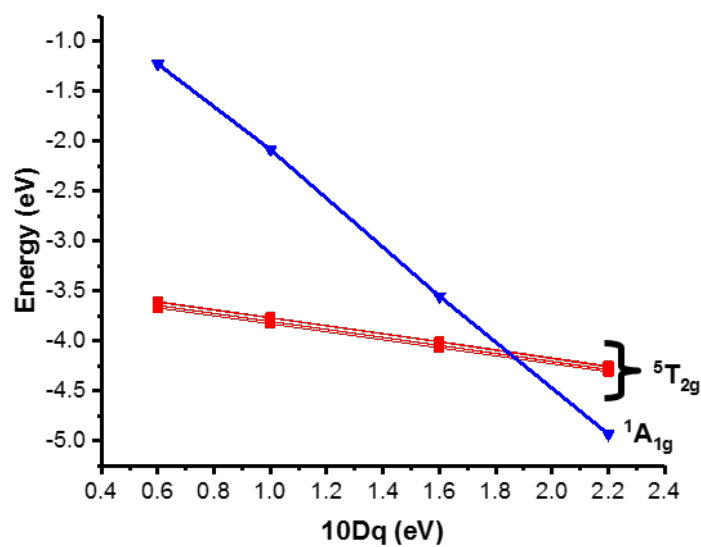

**Figure 8** Energies of the $^1A_{1g}$ eigenfunction and of the spin-orbit split component functions of $^5T_{2g}$ as a function of crystal field strength.

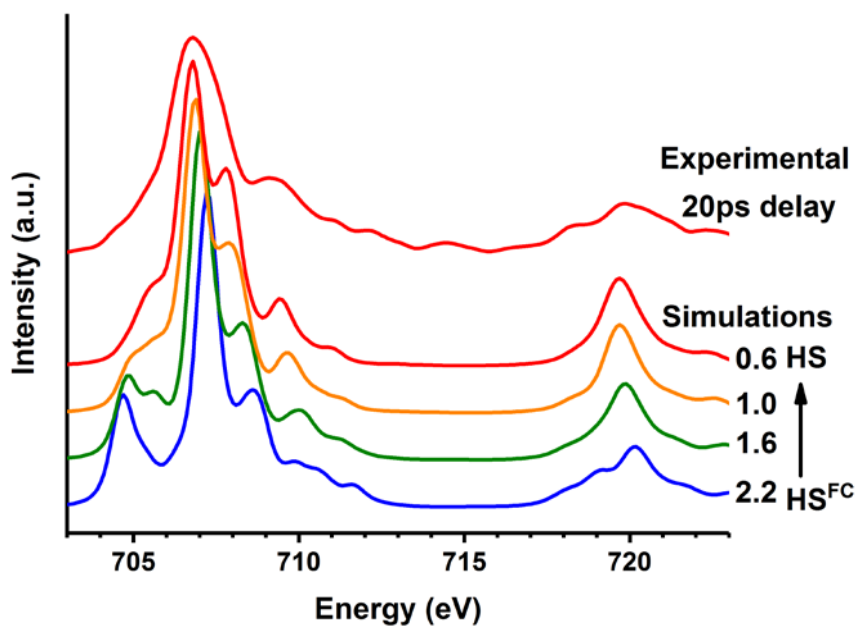

**Figure 9** Top: Experimental excited state $L_{2,3}$-edge spectrum of Fe[Tren(Py)$_3$](PF$_6$)$_2$ 20 ps after pump (Huse *et al.*, 2010). Bottom: LFM simulated $L_{2,3}$-edge spectra of the $^5T_2$ state of an Fe$^{II}$ cation with 10 Dq = 0.6, 1.0, 1.6 and 2.2 eV.

Figure 9 shows the simulated $L_{2,3}$-edge spectra of a quintet species at a range of crystal field strengths, which qualitatively track the relaxation of the quintet Fe$^{II}$ center from the Frank-Condon state just after generation (HS$^{FC}$) to the relaxed state reached after 20 ps (HS). Notably, the isolated peak at 706 eV, which is mainly due to transitions from 2p orbitals to d-orbitals of $t_2$ symmetry, merges into a broad feature at 707-708 eV as the system relaxes. This is because, as 10Dq decreases, the $t_2$ orbitals rise in energy and cause the corresponding absorption feature to move to higher energy.

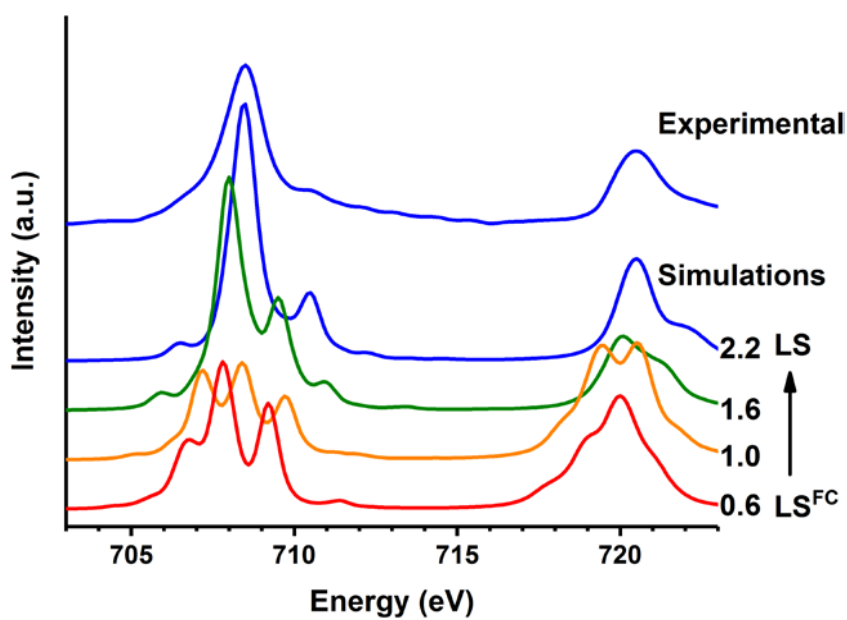

**Figure 10**    $L_{2,3}$-edge spectra of a quintet $Fe^{II}$ cation. Top: Experimental $L_{2,3}$-edge spectrum of Fe[Tren(Py)$_3$](PF$_6$)$_2$ in its ground state (Huse *et al.*, 2010). Bottom: LFM simulated $L_{2,3}$-edge spectra of the $^1A_1$ state of an $Fe^{II}$ cation with 10 Dq = 0.6, 1.0, 1.6 and 2.2 eV.

The differences exhibited by a singlet species when placed in a range of ligand environments are much more dramatic (Figure 10). The $L_3$ features of a Frank-Condon singlet species (LS$^{FC}$) consist of two peaks of comparable intensities and a shoulder at lower energy, whereas almost all intensity in the $L_3$ region of a relaxed singlet species (LS) is concentrated in the single peak at 708 eV.

These examples show that the spectrum of a metal compound may vary significantly as the nuclear geometry relaxes to accommodate a photogenerated excited state. By their very nature, these non-equilibrium states cannot be modeled by reference to the ground state of a model compound. With the finer time resolutions available on new and emerging platforms such as free-electron laser and HHG-based light sources, investigating non-equilibrium photophysics by soft-X-ray spectroscopy will become an increasingly realistic proposition. The flexibility of CTM4XAS in simulating excited state species makes it a good aid for interpreting the spectra of systems in a vibrationally hot state.

## 4. Conclusion

We have shown that simulations of the L- and M-edge absorption spectra of d-d excited states of first-row transition metal complexes can easily be performed using a modified version of the CTM software package CTM4XAS. For two case studies, we show that CTM theory simulations of excited state spectra are in good agreement with experimental data and give additional insights into the systems studied. Furthermore, CTM theory simulations can model the spectra of vibrationally hot species formed at short timescales in ultrafast X-ray spectroscopic experiments.

**Acknowledgements**     This material is based upon work supported by the National Science Foundation under Grant No. 1555245 (to J.V.-W.) and by the William and Janet Lycan Fund at the University of Illinois (to G.S.G.).  We thank Prof. Dr. F. M. F de Groot at Utrecht University for valuable comments. We thank Dr. A. S. Chatterley and Dr. O. Gessner for making available transient data on gaseous ferrocene and helpful discussions.

**References**

Auböck, G. & Chergui, M. (2015). *Nat. Chem.* **7**, 629–633.

Baker, L. R., Jiang, C.-M., Kelly, S. T., Lucas, J. M., Vura-Weis, J., Gilles, M. K., Alivisatos, A. P. & Leone, S. R. (2014). *Nano Lett.* **14**, 5883–5890.

Bressler, C. & Chergui, M. (2004). *Chem. Rev.* **104**, 1781–1812.

Bressler, C. & Chergui, M. (2010). *Annu. Rev. Phys. Chem.* **61**, 263–282.

Butler, P. H. (1981). Point Group Symmetry Applications, Methods and Tables Springer US.

Chatterley, A. S., Lackner, F., Pemmaraju, C. D., Neumark, D. M., Leone, S. R. & Gessner, O. (2016). *J. Phys. Chem. A*. **120**, 9509–9518.

Chen, L. X. (2005). *Annu. Rev. Phys. Chem.* **56**, 221–254.

Chen, L. X., Zhang, X. & Shelby, M. L. (2014). *Chem. Sci.* **5**, 4136–4152.

Cho, H., Strader, M. L., Hong, K., Jamula, L., Gullikson, E. M., Kim, T. K., Groot, F. M. F. de, McCusker, J. K., Schoenlein, R. W. & Huse, N. (2012). *Faraday Discuss.* **157**, 463–474.

Consani, C., Prémont-Schwarz, M., Elnahhas, A., Bressler, C., van Mourik, F., Cannizzo, A. & Chergui, M. (2009). *Angew. Chem. Int. Ed. Engl.* **48**, 7184–7187.

Cowan, R. D. (1981). The Theory of Atomic Structure and Spectra Berkeley: University of California Press.

Fano, U. (1961). *Phys. Rev.* **124**, 1866–1878.

Goulielmakis, E., Loh, Z.-H., Wirth, A., Santra, R., Rohringer, N., Yakovlev, V. S., Zherebtsov, S., Pfeifer, T., Azzeer, A. M., Kling, M. F., Leone, S. R. & Krausz, F. (2010). *Nature*. **466**, 739–743.

Gray, H. B., Sohn, Y. S. & Hendrickson, N. (1971). *J. Am. Chem. Soc.* **93**, 3603–3612.

de Groot, F. (2005). *Coord. Chem. Rev.* **249**, 31–63.

de Groot, F. & Kotani, A. (2008). Core level spectroscopy of solids Boca Raton: CRC Press.

de Groot, F. M. F., Glatzel, P., Bergmann, U., van Aken, P. A., Barrea, R. A., Klemme, S., Hävecker, M., Knop-Gericke, A., Heijboer, W. M. & Weckhuysen, B. M. (2005). *J. Phys. Chem. B*. **109**, 20751–20762.

Hauser, A. (1986). *Chem. Phys. Lett.* **124**, 543–548.

Hauser, A. (2004). *Top. Curr. Chem.* **234**, 155–198.

Hocking, R. K., Wasinger, E. C., de Groot, F. M. F., Hodgson, K. O., Hedman, B. & Solomon, E. I. (2006). *J. Am. Chem. Soc.* **128**, 10442–10451.

Hocking, R. K., Wasinger, E. C., Yan, Y.-L., de Groot, F. M. F., Walker, F. A., Hodgson, K. O., Hedman, B. & Solomon, E. I. (2007). *J. Am. Chem. Soc.* **129**, 113–125.

Hong, K., Cho, H., Schoenlein, R. W., Kim, T. K. & Huse, N. (2015). *Acc. Chem. Res.* **48**, 2957–2966.

Huse, N., Kim, T. K., Jamula, L., McCusker, J. K., de Groot, F. M. F. & Schoenlein, R. W. (2010). *J.*

*Am. Chem. Soc.* **132**, 6809–6816.

Jiang, C.-M., Baker, L. R., Lucas, J. M., Vura-Weis, J., Alivisatos, A. P. & Leone, S. R. (2014). *J. Phys. Chem. C*. **118**, 22774–22784.

Josefsson, I., Kunnus, K., Schreck, S., Föhlisch, A., de Groot, F., Wernet, P. & Odelius, M. (2012). *J. Phys. Chem. Lett.* **3**, 3565–3570.

Laan, G. van der (2006). *Magnetism: A Synchrotron Radiation Approach*, Vol. edited by E. Beaurepaire, H. Bulou, F. Scheurer & J.-P. Kappler, pp. 143–199. Springer Berlin Heidelberg.

Lemke, H. T., Kjær, K. S., Hartsock, R., van Driel, T. B., Chollet, M., Glownia, J. M., Song, S., Zhu, D., Pace, E., Matar, S. F., Nielsen, M. M., Benfatto, M., Gaffney, K. J., Collet, E. & Cammarata, M. (2017). *Nat. Commun.* **8**, 15342.

Milne, C. J., Penfold, T. J. & Chergui, M. (2014). *Coord. Chem. Rev.* **277–278**, 44–68.

Monat, J. E. & McCusker, J. K. (2000). *J. Am. Chem. Soc.* **122**, 4092–4097.

Okada, K. & Kotani, A. (1993). *J. Electron Spectros. Relat. Phenomena*. **62**, 131–140.

Otero, E., Kosugi, N. & Urquhart, S. G. (2009). *J. Chem. Phys.* **131**, 114313.

Piepho, S. B. & Schatz, P. N. (1983). Group Theory in Spectroscopy with Applications to Magnetic Circular Dichroism New York: John Wiley & Sons, Inc.

Roemelt, M., Maganas, D., DeBeer, S. & Neese, F. (2013). *J. Chem. Phys.* **138**, 204101.

Stavitski, E. & de Groot, F. M. F. (2010). *Micron*. **41**, 687–694.

Thole, B. T., van der Laan, G., Fuggle, J. C., Sawatzky, G. A., Karnatak, R. C. & Esteva, J.-M. (1985). *Phys. Rev. B*. **32**, 5107–5118.

van der Veen, R. M., Penfold, T. J. & Zewail, A. H. (2015). *Struct. Dyn.* **2**, 24302.

Vura-Weis, J., Jiang, C.-M., Liu, C., Gao, H., Lucas, J. M., de Groot, F. M. F., Yang, P., Alivisatos, A. P. & Leone, S. R. (2013). *J. Phys. Chem. Lett.* **4**, 3667–3671.

Zewail, A. H. (2000). *J. Phys. Chem. A*. **104**, 5660–5694.

Zhang, K., Lin, M.-F., Ryland, E. S., Verkamp, M. A., Benke, K., de Groot, F. M. F., Girolami, G. S. & Vura-Weis, J. (2016). *J. Phys. Chem. Lett.* **7**, 3383–3387.

Zhang, W., Alonso-Mori, R., Bergmann, U., Bressler, C., Chollet, M., Galler, A., Gawelda, W., Hadt, R. G., Hartsock, R. W., Kroll, T., Kjær, K. S., Kubiček, K., Lemke, H. T., Liang, H. W., Meyer, D. A., Nielsen, M. M., Purser, C., Robinson, J. S., Solomon, E. I., Sun, Z., Sokaras, D., van Driel, T. B., Vankó, G., Weng, T.-C., Zhu, D. & Gaffney, K. J. (2014). *Nature*. **509**, 345–348.